\documentclass[conference]{IEEEtran}
\usepackage{graphicx} 
\usepackage{subfigure}
\usepackage[noadjust]{cite}
\usepackage{amsmath,amssymb, amsfonts}

\usepackage{fancyhdr}
\newcommand{\defeq}{\stackrel{\Delta}{=}}
\newcommand{\gf}[1]{\text{GF}(#1)}
\newcommand{\Zbb}{\mathbb{Z}}

\newcommand{\alphabet}{\mathcal{U}}
\newcommand{\symb}{u}

\usepackage{caption}
\usepackage{tikz}
\usetikzlibrary{arrows.meta}
\usetikzlibrary{shapes.geometric,plotmarks,backgrounds,fit,calc,circuits.ee.IEC}
\usetikzlibrary{decorations.pathreplacing}
\DeclareMathOperator*{\vsargmax}{argmax}

\IEEEoverridecommandlockouts

\begin{document}
%
\title{Non-Binary Polar Codes for Spread-Spectrum Modulations}



\author{Valentin Savin, CEA-LETI, Universit\'e Grenoble Alpes,  France (valentin.savin@cea.fr)%
\thanks{This work was partially supported by the French Agence Nationale de la Recherche (ANR), under grant number ANR-19-CE25-0013 (QCSP project).}}


%


\maketitle

\thispagestyle{fancy}

\begin{abstract}
This paper proposes a new coded modulation scheme for reliable transmission of short data packets at very low signal-to-noise ratio, combining cyclic code shift keying modulation and non-binary polar coding. We consider non-binary polar codes defined over Galois fields, and propose a new design methodology, aimed at optimizing the choice of the kernel coefficients. Numerical results show that the system performance is close to the achievable limits in the finite blocklength regime.
\end{abstract}


%

\section{Introduction}


Recent years have seen an explosive growth in the number of devices connected and controlled by the Internet.
The wide range of applications for Internet of Things (IoT) technology is usually divided into two use-cases, known as either ``critical IoT'' or ``massive IoT''. The latter is  characterized by a high density of connected devices, small data payloads, and low sensitivity levels, due to stringent constraints on the device energy consumption and cost. Maximizing the spectral efficiency of an IoT network is a key prerequisite for providing massive connectivity. Yet, the first wave of IoT standards are far from achieving the  spectral efficiency targets. They implement sub-optimal forward error correction schemes, such as convolutional or Turbo codes combined with repetition codes (EC-GSM, Narrowband-IoT and LTE-M), simple Hamming codes (LoRa), or simply omit any FEC capability (SigFox). For instance, in Narrowband-IoT, the coded block may be repeated up to 128 times in uplink mode, and up to 512 times in downlink. While the main advantage of repetition coding is the ease of implementation, it has poor error correction performance and does not improve the energy efficiency since it provides no coding gain.

In this paper we investigate an alternative strategy to achieve low  levels of sensitivity with increased spectral efficiency, based on advanced channel coding, combined with Cyclic Code Shift Keying (CCSK) modulation. The CCSK modulation is  a direct-sequence spread-spectrum technique, which has been shown to provide significant advantages in terms of both demodulation and synchronization, when combined with non-binary channel coding \cite{abassi2013non,saied2020quasi}. Accordingly, in this work we consider the use of non-binary polar codes~\cite{sasouglu2009polarization,mori2010channel, mori2010non, sasoglu2011polar,chiu2014non,yuan2018construction}, as channel coding technique.  
They may provide significant coding gain, thus enabling transmission at very low power.  However, to exploit their full potential  non-binary polar codes have to be carefully optimized, which is even more true for small data payloads. 

The paper is organized as follows. In Section~\ref{sec:system-model}, we introduce the system model and derive the achievable rates in the both asymptotic and finite blocklength regimes. In Section~\ref{sec:nb-polar-coding}, we shortly discusses non-binary polar coding, and present non-binary polar codes defined over Galois fields. In Section~\ref{sec:design-method-polar}, we present the non-binary code design methodology, aimed at optimizing the choice of the kernel coefficients. Numerical results are presented in Section~\ref{sec:simulation-results-polar}.


\section{System Model}
\label{sec:system-model}

\subsection{CCSK Modulation}
\label{subsec:ccsk_modulation}
We denote by $\alphabet \defeq \{0, 1, \dots, q-1\}$ the set of integers comprised between $0$ and $q-1$, where $q = 2^p$ is a power of $2$. We shall further identify $\alphabet \cong \Zbb_2^p \defeq \{0, 1\}^p$, by identifying an integer to its binary representation, $\symb  \ \cong \ (\symb(0),\dots,\symb(p-1))$.
%
Let $\mathbf{P}_0 \defeq \left( \mathbf{P}_0(0), \mathbf{P}_0(1), \dots ,\mathbf{P}_0(q-1) \right)$ be a pseudo-random noise (PN) sequence, of length $q$, with good cross-correlation properties ({\em e.g.}, $\mathbf{P}_0$ may be generated  by  a linear feedback shift register, with primitive feedback polynomial). We assume that $\mathbf{P}_0(i) \in \{-1,+1\}, \forall i=0,\dots,q-1$.
For $\symb\in \alphabet$, we define $\mathbf{P}_\symb$ to be the sequence obtained by   shifting  $\mathbf{P}_0$ circularly to the left, by $\symb$ positions, that is 
\begin{equation}
\mathbf{P}_\symb(i) \defeq \mathbf{P}_0(i+\symb\mod q), \ \forall i=0,\dots,q-1.
\end{equation}
The CCSK modulation maps an element $\symb\in \alphabet$ to the sequence $\mathbf{P}_\symb$.
%
The ratio $p/q$ is referred to as the {\em spreading factor} of the modulation.



\subsection{Demodulation}
\noindent We will use the following notation. 
\begin{itemize}
\item $U= (U(0),\dots,U(p-1))$ denotes a uniform random variable, with values in $\alphabet$. Realizations of a $U$ represent unmodulated symbols (input to the CCSK modulation). 
\item $X = (X(0),X(1),\dots,X(q-1)) \in \{-1, +1\}^q$ denotes the random variable defined by modulating $U$. Hence, $X = \mathbf{P}_\symb \Leftrightarrow U = \symb$.
\item  $Y = (Y(0),Y(1),\dots,Y(q-1)) \in \mathbb{R}^q$ denotes the received signal. 
\item $\tilde{Y} = (\tilde{Y}(0),\tilde{Y}(1),\dots,\tilde{Y}(q-1)) \in \mathbb{R}^q$, where 
\begin{equation}
\tilde{Y}(i) \defeq \log \frac{\Pr\left(X(i)=+1 \mid Y(i)\right)}{\Pr\left(X(i)=-1 \mid Y(i)\right)} 
\end{equation}
\end{itemize}
Assuming that the CCSK modulated signal $X$  undergoes real additive white Gaussian noise, we have
\begin{equation}
Y(i) = X(i) + Z(i) \ \text{ and } \ \tilde{Y}(i)= \frac{2}{\sigma^2}Y(i),  
\end{equation}
where $Z(i)$ are real-valued, mutually independent, normal random variables, with mean $0$ and variance $\sigma^2$.

Given the received signal $Y$,  the symbol-level Log-Likelihood Ratio (LLR) values are defined by
\begin{equation}\label{eq:defGamma}
\Gamma(\symb) \defeq \log \frac{\Pr\left(U=0 \mid Y\right)}{\Pr\left(U=\symb \mid Y\right)}, \ \ \forall \symb\in \alphabet
\end{equation}
Hence, we have
\begin{align}
\Gamma(\symb) & = \log \frac{\Pr\left(U=0 \mid Y\right)}{\Pr\left(U=\symb \mid Y\right)} 
  = \log \frac{\Pr\left(X=\mathbf{P}_0 \mid Y\right)}{\Pr\left(X = \mathbf{P}_\symb \mid Y\right)} \\
   &= \sum_{i=0}^{q-1} \log \frac{\Pr\left(X(i)=\mathbf{P}_0(i) \mid Y(i)\right)}{\Pr\left(X(i)=\mathbf{P}_\symb(i) \mid Y(i)\right)} \\
  & = \sum_{i=0}^{q-1} \frac{\mathbf{P}_0(i)-\mathbf{P}_\symb(i)}{2}\log  \frac{\Pr\left(X(i)=+1 \mid Y(i)\right)}{\Pr\left(X(i)=-1 \mid Y(i)\right)} \\
  &= \frac{1}{2} \left( \tilde{Y} \cdot\mathbf{P}_0 - \tilde{Y} \cdot \mathbf{P}_\symb \right) 
\end{align}
where $\tilde{Y} \cdot \mathbf{P} \defeq \sum_i \tilde{Y}(i)\mathbf{P}(i)$ denotes the usual dot product of sequences (vectors) $\tilde{Y}$ and $\mathbf{P}$. Since that $\mathbf{P}_\symb$ is a circular shifted version of $\mathbf{P}_0$, dot products $Y\cdot \mathbf{P}_\symb$, $\symb\in\alphabet$, can be conveniently computed by using the discrete Fourier transform, denoted by ${\cal F}$. 
%
%
Precisely, 
\begin{equation}
Y \cdot \mathbf{P}_\symb = (Y\ast\mathbf{P}_0) (\symb) = {\cal F}^{-1}\left( {\cal F}(Y)^{*}\cdot {\cal F}(\mathbf{P}_0) \right) (\symb),
\end{equation}
where ${\cal F}(Y)^{*}$ is the complex conjugate of ${\cal F}(Y)$.
%
%
%
%
%
%
%
 Finally, from the above LLR values, the probability distribution of $U$ conditional on $Y$ can be computed by
\begin{equation}\label{eq:proba_U}
\Pi(\symb) \defeq \Pr(U=\symb \mid Y) = \frac{e^{-\Gamma(\symb)}}{\sum_{\symb^\prime\in \alphabet} e^{-\Gamma(\symb^\prime)}}
\end{equation}

In case that the unmodulated symbols are encoded by a non-binary code, the received signal is first demodulated, then the symbol-level LLR values (or equivalently, the corresponding probability distribution on the alphabet ${\cal U}$) are supplied to the non-binary decoder. 

%

\subsection{Achievable Rates}
We assume that the unmodulated symbols are encoded by a non-binary code, with alphabet $\alphabet$. The {\em  coding rate}  is the ratio between the number of source symbols and the total number of encoded symbols.

\subsubsection*{Asymptotic Blocklength Regime}
By Shannon's noisy-channel coding theorem \cite{shannon1948mathematical}, the {\em maximum achievable (coding) rate}, denoted in the sequel by $R$,  is given by mutual information between the input $U$ of the CCSK modulation and the output $Y$ of the channel
\begin{equation}\label{eq:actualR}
R \defeq I(U ; Y) = H(U) - H(U | Y), 
\end{equation} 
where $H$ denotes the Shannon  entropy. We assume a base-$q$ logarithm for the entropy, such that $R \in [0, 1]$. Since the channel is symmetric, its capacity is achieved for an uniformly distribution input $U$. Hence, we have $H(U)=1$, while the conditional entropy $H(U | Y)$ can be conveniently estimated numerically, by averaging over the channel output $Y$,
\begin{equation}
 H(U | Y) = \mathbb{E}_Y \big[-\sum_{\symb\in \alphabet} \Pi(\symb) \log_q\Pi(\symb) \big].
\end{equation}


\subsubsection*{Finite Blocklength Regime}
In the non-asymptotic regime, the backoff from channel capacity can be accurately characterized by a parameter known as {\em channel dispersion}~\cite{polyanskiy2010channel}. 
Specifically, the maximum achievable coding rate can be tightly approximated by
\begin{equation}\label{eq:normal_approx}
R^{\ast} \defeq R - \sqrt{\frac{V}{N}}Q^{-1}(\varepsilon),
\end{equation}
where $R$ is the channel capacity, 
and $V$ is the channel dispersion. 
$R^{\ast}$ is usually referred to as the {\em normal approximation}.
%
%
 Using \cite[Theorem 49]{polyanskiy2010channel},  the channel dispersion parameter can be computed as
\begin{align}
V &= H_2(U \mid Y) - H(U \mid Y)^2, \\
\text{where }  H_2(U \mid Y) &\defeq  \textstyle \mathbb{E}_Y \big[-\sum_{\symb\in \alphabet} \Pi(\symb) \log_q^2\Pi(\symb) \big],
\end{align}
which can again be be conveniently estimated numerically by Monte-Carlo simulation.

\section{Non-Binary Polar Codes} 
\label{sec:nb-polar-coding}

Two main approaches have been proposed in the literature for polarizing channels with non-binary input alphabets. The first one relies on using  higher-dimensional non-binary kernels, that is, kernels of size $\ell \times \ell$, with $\ell > 2$ \cite{mori2010channel, mori2010non, sasoglu2011polar}. Such an approach is characterized by an increased complexity, due to both the size of the non-binary alphabet, and the higher kernel dimension. A different approach, proposed in \cite{sasouglu2009polarization}, is to use a randomized construction, based on the original kernel proposed by Arikan. Precisely, the kernel transformation is defined by $(u_0,u_1) \mapsto (u_0\oplus u_1, \pi(u_1))$, 
where $\pi$ is a random permutation of the non-binary alphabet (here `$\oplus$' may be any additive group operation on the non-binary alphabet). 
Channel polarization essentially states that for a random choice of permutations throughout the recursive channel combining and splitting procedure, the synthesized virtual channels  polarize to either useless or perfect channels.  
In this case, the {\em polar code construction} encompasses  the {\em choice} of  both  channel combining permutations and  virtual channels used to transmit information symbols.  
%
%
%
Of course, once the code is constructed, randomness does no longer exist, and the complexity of polar code encoding and decoding is essentially the same as for the Arikan's kernel.

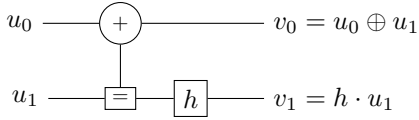
\begin{figure}[!t]
\centering
    \begin{tikzpicture}
        \draw 
            (0,0) node[circle, draw, scale=.8] (X) {$+$}
            (X) ++(0, -1) node[draw, scale=.8] (E) {$=$}
            (X.south) to (E.north) 
            (E.east) to ++(0.5, 0) node[draw, right, scale=1] (h) {$h$}
            (X.west) to ++(-0.75, 0) node[left] {$u_0$}
            (E.west) to ++(-0.75, 0) node[left] {$u_1$}
            (h.east) to ++(0.75, 0) node[right] {$v_1 = h\cdot u_1$}
            (X.east) to ($(h.east)+(0.75,1)$) node[right] (x1) {$v_0 = u_0 \oplus u_1$}
            ;

    \end{tikzpicture}
\caption{GF kernel, with $h$ a non-zero GF-element.}
\label{fig:gf-kernel}
\end{figure}

The non-binary polar codes considered in this work are based on the randomized construction described above. However, we consider non-binary polar codes defined over Galois fields (GF), and rather than random GF permutations, we consider linear permutations defined by the multiplication with a non-zero GF element~\cite{chiu2014non,yuan2018construction}. Precisely, using the notation from the previous section, we denote by $W(Y \mid U)$ the channel with non-binary input alphabet ${\cal U}$, encompassing both the CCSK modulation and the actual transmission channel. We further endow ${\cal U}$ with a GF structure, with (additive, multiplicative) operations denoted  by $(\oplus, \cdot)$. Finally, the kernel transformation, illustrated in Fig.~\ref{fig:gf-kernel}, is defined by $(u_0,u_1) \mapsto (v_0, v_1) \defeq (u_0\oplus u_1, h\cdot u_1)$, where $h\in {\cal U}^*$ (the multiplicative group of non-zero GF elements), referred to as {\em kernel coefficient}.

 
\section{Design Methodology} 
\label{sec:design-method-polar}

Throughout the rest of the paper, we denote by $\gf{q}$ the Galois field with $q$ elements, and further identify $\alphabet \cong \gf{q}$. 

\subsection{Optimization of the kernel coefficients}
\label{subsec:optimization-procedure}
While the polarization result in~\cite{sasouglu2009polarization,chiu2014non} essentially states that a random choice of the kernel coefficients is good enough, it might not be optimal.  Thus,  the optimization of the kernel coefficients is aimed at accelerating the speed of polarization of the synthesized virtual channels. There are three parameters that may be used to describe the polarization process: the mutual information, the Bhattacharyya parameter, and the error probability of the synthesized virtual channels. The former approaches $0$ (respectively\footnote{We assume here that the mutual information is normalized (expressed in terms of symbols per channel use), thus taking values between $0$ and $1$.}, $1$) if and only if the latter two approach $1$ (respectively, $0$). Any of these parameters may be used within the proposed optimization procedure, and for the moment we shall simply use {\em polarizing parameter} to refer to any of them. 
To accelerate the speed of polarization, we choose the kernel coefficients so as to maximize the difference between the polarizing parameters of the bad and good channels synthesized by the channel combining and splitting procedure. 

\begin{figure}[!t]
    \includegraphics[width=\linewidth]{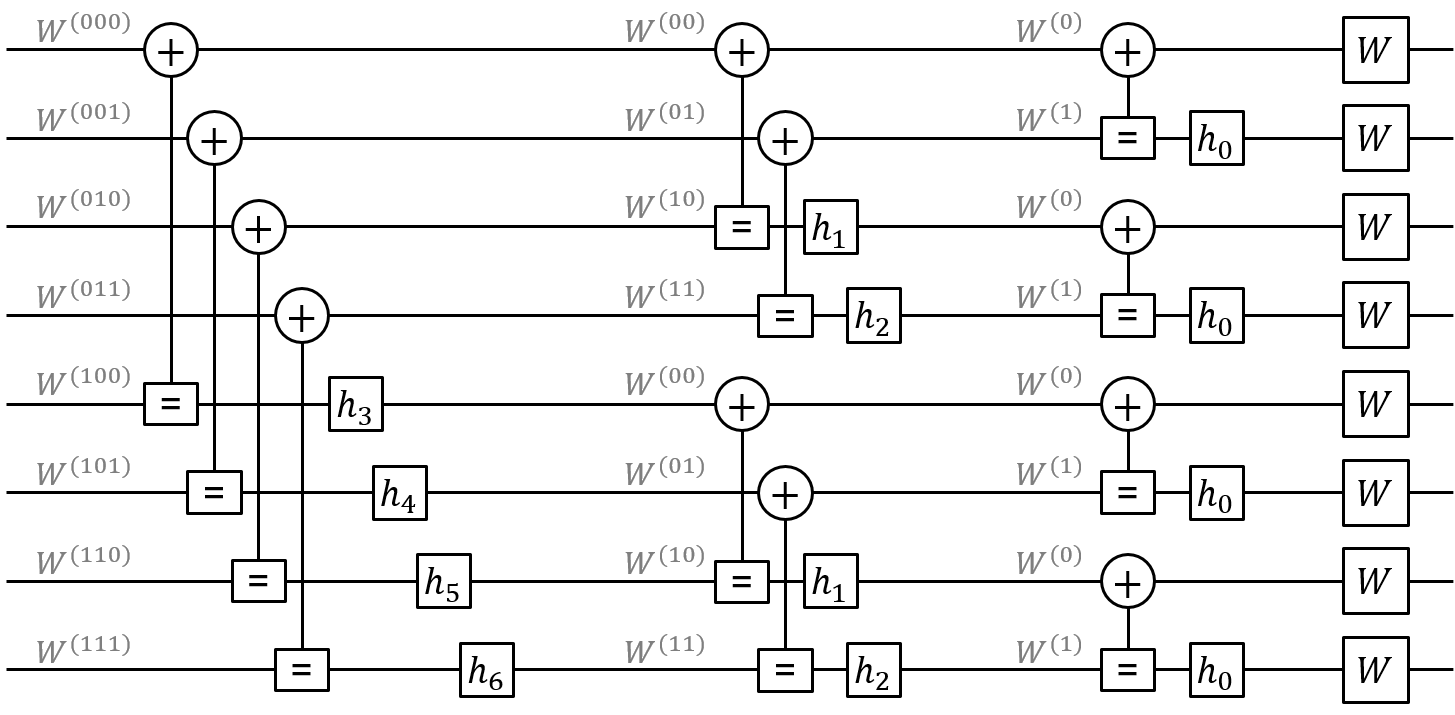}\\
    {\small \hspace*{12mm}  3rd  step \hspace*{18mm} 2nd  step \hspace*{10mm} 1st  step }
    \caption{Non-binary polar code of length $N=8$, corresponding to $n=3$ steps of polarization}
    \label{fig:polar-optimization}
\end{figure}

\smallskip The optimization procedure is illustrated at Fig.~\ref{fig:polar-optimization}, for a polar code of length $N=8$, corresponding to $n=3$ {\em polarization steps}. The original non-binary channel is denoted by $W$. We denote by $W^{(0)}$ and $W^{(1)}$ the bad and good channels, respectively, after one step of polarization. Then, for $n > 0$, we define recursively 
\begin{equation}
W^{(i_1\dots i_n)} := \left(W^{(i_1\dots i_{n-1})}\right)\,\!\!^{(i_n)},\ \forall (i_1\dots i_n) \in \{0,1\}^n
\end{equation}
In Fig.~\ref{fig:polar-optimization}, we have indicated on each horizontal wire the virtual channel $W^{(i_1i_2\dots)}$ ``seen'' by the corresponding symbol throughout the polarization process.  All the kernels on the first (right-most) polarization step  combine two copies of the $W$ channel. Therefore, only one coefficient needs to be optimized, denoted by $h_0$. We define $h_0$ as
\begin{equation}
h_0 := \vsargmax_{h\in\gf{q}} \left| P^{(0)}(h) -  P^{(1)}(h) \right|,
\end{equation}
where $P^{(0)}(h)$ and $P^{(1)}(h)$ denote the polarizing parameters of $W^{(0)}$ and $W^{(1)}$ channels, respectively, assuming that the channel combining coefficient is equal to $h$. We numerically estimate the values of $P^{(0)}(h)$ and $P^{(1)}(h)$, for all $h\in \gf{q}\setminus\{0\}$, based on Monte Carlo simulation (see also Section~\ref{subsec:polar-code-construction}).

 Once the value of $h_0$ is determined, we can optimize the kernel coefficients for the second (middle) polarization step. There are two different types of kernels on the second polarization step,  combining either two copies of $W^{(0)}$, or two copies of $W^{(1)}$. Therefore, two coefficients need to be optimized, denoted by $h_1$ and $h_2$ in Fig.~\ref{fig:polar-optimization}. Hence, we define
\begin{align}
h_1 := \vsargmax_{h\in\gf{q}} \left| P^{(00)}(h) -  P^{(10)}(h) \right|, \\
h_2 := \vsargmax_{h\in\gf{q}} \left| P^{(10)}(h) -  P^{(11)}(h) \right|,
\end{align}
where $P^{(i_1i_2)}(h)$ denotes the polarizing parameters of the $W^{(i_1i_2)}$ channel, assuming the channel combining coefficient on the second polarization step is equal to  $h$. The value of  $P^{(i_1i_2)}(h)$ is again estimated numerically through Monte Carlo simulation. Then the optimization process continue recursively, until the desired number of polarization steps is reached.

\subsection{Non-binary polar decoding}
\label{subsec:decoding-polar}
\begin{figure}[!t]
\centering
\subfigure[Bad channel decoding]{%
    \begin{tikzpicture}[scale=0.75]
        \draw
            (0,0) node[circle, draw, scale=.8] (X) {$+$}
            (X) ++(0, -1) node[draw, scale=.8] (E) {$=$}
            (X.south) to (E.north) 
            ; 
        \draw[->,>=Latex]
            (X.west) to ++(-0.75, 0) node[left] {$\Pi_U^{(0)}$}
            ;
        \draw[<-,>=Latex]
            (E.east) to ($(E.east)+(0.5,0)$)  node[draw, right] (h) {$h$}
            ;
        \draw[<-,>=Latex]
            (X.east) to ($(h.east)+(0.75,1)$) node[right] (x1) {$\Pi_V^{(0)}$}
            ;
        \draw
            (h.east) to ++(0.75, 0) node[right] {$\Pi_V^{(1)}$}
            ;
    \end{tikzpicture}
    \label{fig:bad-channel-decoding}
}\hfill%
\subfigure[Bad channel decoding]{%
    \begin{tikzpicture}[scale=0.75]
        \draw
            (0,0) node[circle, draw, scale=.8] (X) {$+$}
            (X) ++(0, -1) node[draw, scale=.8] (E) {$=$}
            (X.south) to (E.north) 
            ;
        \draw[<-,>=Latex]    
            (X.west) to ++(-0.75, 0) node[left] {$u_0$}
            ;
        \draw[->,>=Latex]
            (E.west) to ++(-0.75, 0) node[left] {$\Pi_U^{(1)}$}
            ;
        \draw[<-,>=Latex]
            (E.east) to ($(E.east)+(0.5,0)$)  node[draw, right] (h) {$h$}
            ;
        \draw[<-,>=Latex]
            (X.east) to ($(h.east)+(0.75,1)$) node[right] (x1) {$\Pi_V^{(0)}$}
            ;
        \draw
            (h.east) to ++(0.75, 0) node[right] {$\Pi_V^{(1)}$}
            ;
    \end{tikzpicture}
    \label{fig:good-channel-decoding}
}
\caption{Decoding of bad and good virtual channels. For decoding the good channel, the decoder uses the knowledge of $u_0$ (genie decoder), or an estimate of it, $\hat{u}_0$ (real decoder).}
\label{fig:decoding-rules}
\end{figure}
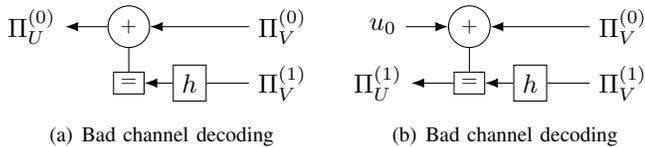

We first consider the decoding of a non-binary kernel, which is illustrated at Fig.~\ref{fig:decoding-rules}. As before, let $u_0,u_1\in \gf{q}$ denote the kernel inputs, and  $v_0,v_1\in\gf{q}$ denote the kernel outputs. Decoding operates in the opposite direction, {\em i.e.}, it takes as inputs $\Pi_V^{(0)}$ and $\Pi_V^{(1)}$, the probability distribution  functions (PDFs) of $v_0$ and $v_1$, respectively, and outputs $\Pi_U^{(0)}$ and $\Pi_U^{(1)}$, the PDFs of $u_0$ and $u_1$, respectively. It can be easily seen that $\Pi_U^{(0)}$ and $\Pi_U^{(1)}$ can be computed from $\Pi_V^{(0)}$ and $\Pi_V^{(1)}$, by the following formulas:
\begin{align}
\Pi_U^{(0)}(u) &= \sum_{u'\in\gf{q}} \Pi_V^{(0)}(u\oplus u') \Pi_V^{(1)}(h\cdot u')\\
\Pi_U^{(1)}(u) &= \eta \Pi_V^{(0)}(u_0\oplus u) \Pi_V^{(1)}(h\cdot u),\label{eq:good-channel-decoding-ga}
\end{align}
where $\eta$ is a normalization factor, determined such that $\sum_{u\in\gf{q}} \Pi_U^{(1)}(u) = 1$.
In equation~(\ref{eq:good-channel-decoding-ga}), the computation of $\Pi_U^{(1)}(u)$ requires the knowledge of  $u_0$. Such a decoder is referred to as {\em genie-aided}, and it is used at the code design stage. For a  {\em real-world decoder}, used to decode a codeword transmitted over a noisy channel, equation~(\ref{eq:good-channel-decoding-ga}) is replaced by 
\begin{align}
\Pi_U^{(1)}(u) &= \eta \Pi_V^{(0)}(\hat{u}_0\oplus u) \Pi_V^{(1)}(h\cdot u),\label{eq:good-channel-decoding}
\end{align}
where either $\hat{u}_0 =u_0$ if the latter is known (frozen bad channel), or $\hat{u}_0 = \vsargmax_{u\in\gf{q}} \Pi_U^{(0)}(u)$ is the estimate of $u_0$, otherwise.

The successive cancellation (SC) decoder (either genie-aided or real-world) uses the above kernel decoding rules in a recursive manner, so as to propagate the PDFs of the transmitted symbols from the right-hand side (transmission channel side) to the left-hand side of the polar graph. There is one such a recursion for each $W^{(i_1\dots i_n)}$ channel (deriving from the recursive definition of the channel), which are then 
decoded successively.

\subsection{Choice of the polarizing parameter and code construction}
\label{subsec:polar-code-construction}
%
Since we are interested on the error rate performance of the constructed polar code, we take the polarizing parameter used within the optimization procedure (from Section~\ref{subsec:optimization-procedure}) to be the error probability of the synthesized virtual channels. An efficient way to numerically estimate the error probability of the synthesized virtual channels is described below, where $u^{(i_1\dots i_n)}\in\gf{q}$ denotes the input of the $W^{(i_1\dots i_n)}$ virtual channel, $(i_1\dots i_n)\in \{0,1\}^n$. 
%
\begin{list}{}{\setlength{\leftmargin}{5mm}\setlength{\itemindent}{0mm}\setlength{\labelwidth}{5mm}}
\item[1)] Randomly generate a set of inputs $\{u^{(i_1\dots i_n)} : (i_1\dots i_n)\in \{0,1\}^n\}$, encode them, and transmit the obtained codeword over the non-binary channel.
\item[2)] Run the genie-aided SC decoder to determine the PDFs of the virtual channels' inputs $u^{(i_1\dots i_n)}$, denoted by $\Pi_U^{(i_1\dots i_n)}$.
\item[3)] Hence, the {\em one-run error probability} of the virtual channel $W^{(i_1\dots i_n)}$ is given by
$P^{(i_1\dots i_n)}_{\text{one-run}} = 1 - \Pi_U^{(i_1\dots i_n)}\left( u^{(i_1\dots i_n)} \right)$.
%
\item[$\circlearrowright$] Repeating the steps 1--3 many times,  the error probability of the virtual channel $W^{(i_1\dots i_n)}$ is estimated by taking the average of the one-run error probability:
\begin{equation}
P^{(i_1\dots i_n)} = \mathbf{E}\left[ P^{(i_1\dots i_n)}_{\text{one-run}} \right]
\end{equation}
\end{list}

 The above procedure is used recursively within the optimization procedure from Section~\ref{subsec:optimization-procedure},  to optimize the kernel coefficients at the different polarization steps. Moreover, once the optimization procedure completed, we may use the $P^{(i_1\dots i_n)}$ values to sort the virtual channels from the best (lowest error probability) to the worst (higher error probability) one, and then use the  best channels 
 to transmit information symbols.  This completes the polar code construction, as we have made a choice of the kernel coefficients, and determined the virtual channels to use for transmitting information symbols. 
 Moreover, when  the polar code construction completes, we also get an estimate the SC decoding error probability, denoted $\overline{\text{WER}}_\text{SC}$. To simplify the notation, let us denote by $P^{(1)},\dots,P^{(K)}$ the error probability of the $K$ virtual channels carrying information symbols. Then, we have
 \begin{equation}
 \textstyle
 \overline{\text{WER}}_\text{SC} := 1 - \prod_{k=1,\dots,K} \left(1 - P^{(k)} \right)
 \end{equation}
 For binary polar codes, $\overline{\text{WER}}_\text{SC}$ is known to provide a tight upper-bound on the word error rate (WER) performance of the SC decoder (which also explains the notation). In Section~\ref{sec:simulation-results-polar}, we will show that the same is true for non-binary polar codes. 
 

 \section{Numerical Results} 
\label{sec:simulation-results-polar}

\begin{figure*}[!t]
\centering
\subfigure[Polar codes defined on $\gf{64}$]{\includegraphics[width=.33\linewidth]{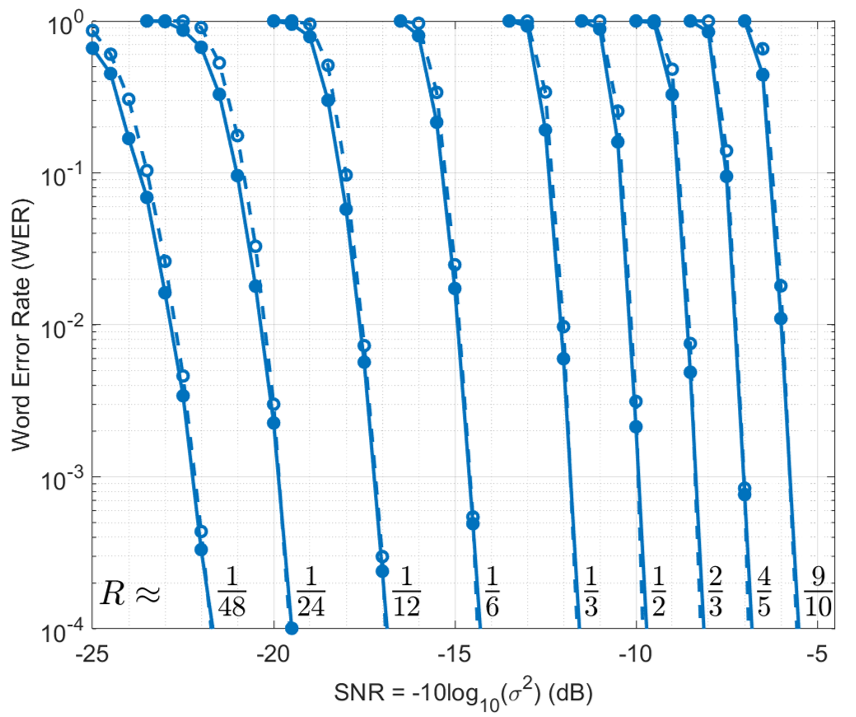}\label{fig:wer_polar_ccsk_p6}}\hfill%
\subfigure[Polar codes defined on $\gf{256}$]{\includegraphics[width=.33\linewidth]{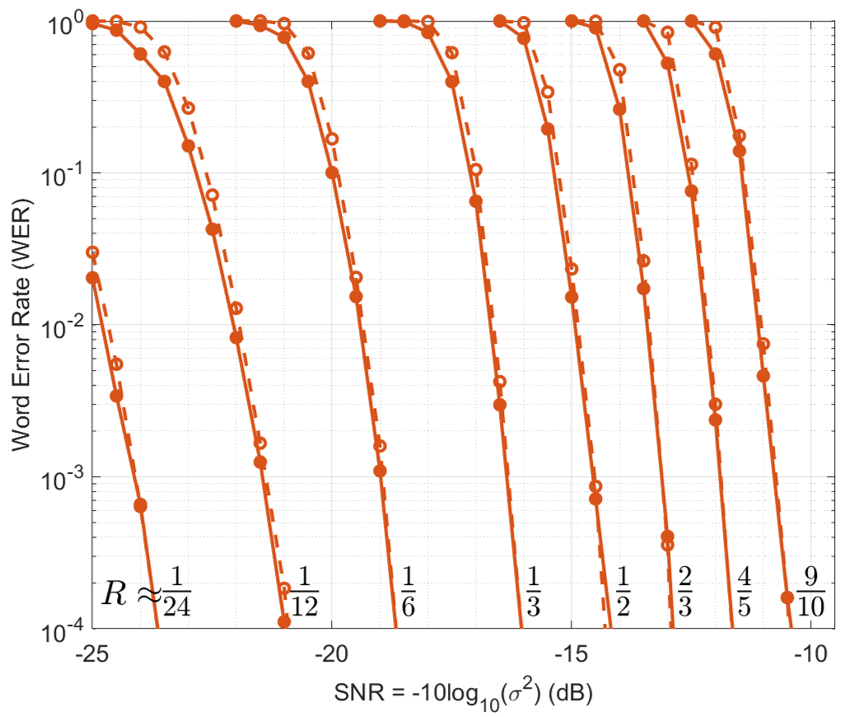}\label{fig:wer_polar_ccsk_p8}}\hfill%
\subfigure[Polar codes defined on $\gf{1024}$]{\includegraphics[width=.33\linewidth]{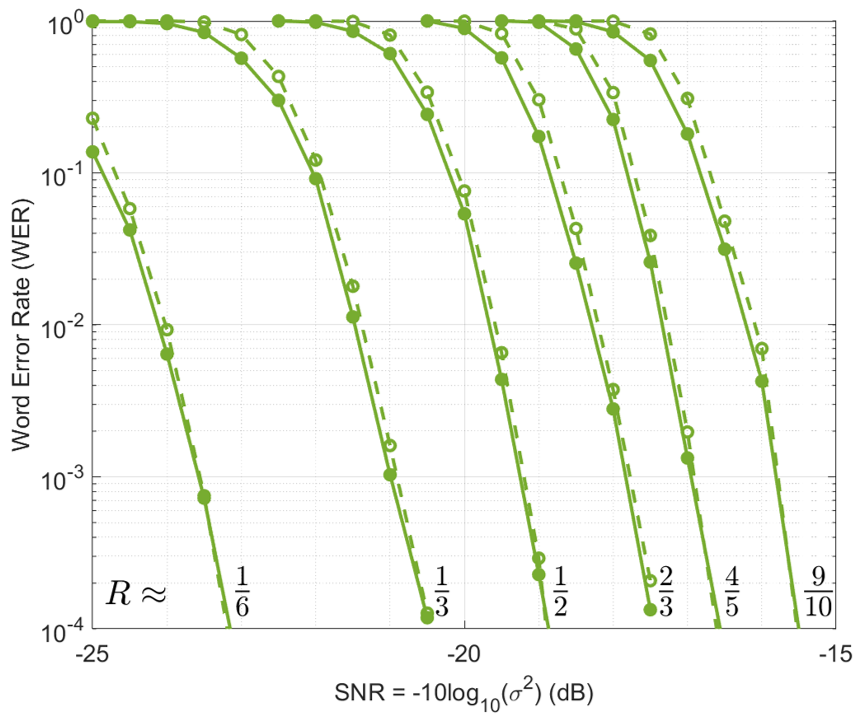}
\label{fig:wer_polar_ccsk_p10}}
\caption{WER performance for various native coding rate values $R$. Solid curves correspond to Monte Carlo simulation results, while dashed curves show the WER estimated at the code construction stage.}
\label{fig:wer_polar_ccsk}
\end{figure*}

\begin{figure*}[!t]
\subfigure[Achievable native coding rates (linear scale)]{\includegraphics[width=.33\linewidth]{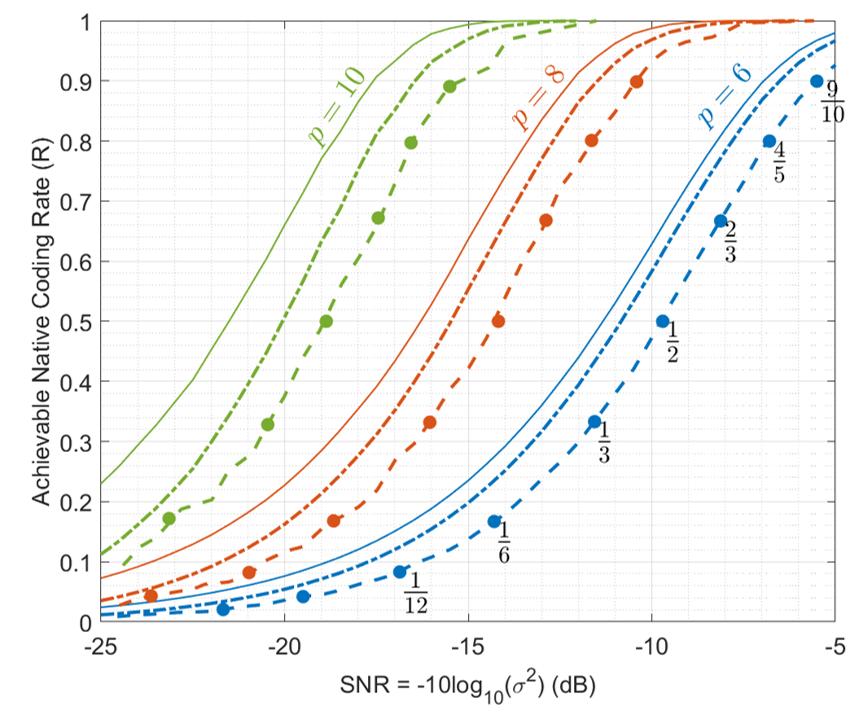}\label{fig:ccsk_achieve_native_rate_lin}}\hfill%
\subfigure[Achievable native coding rates (log scale)]{\includegraphics[width=.33\linewidth]{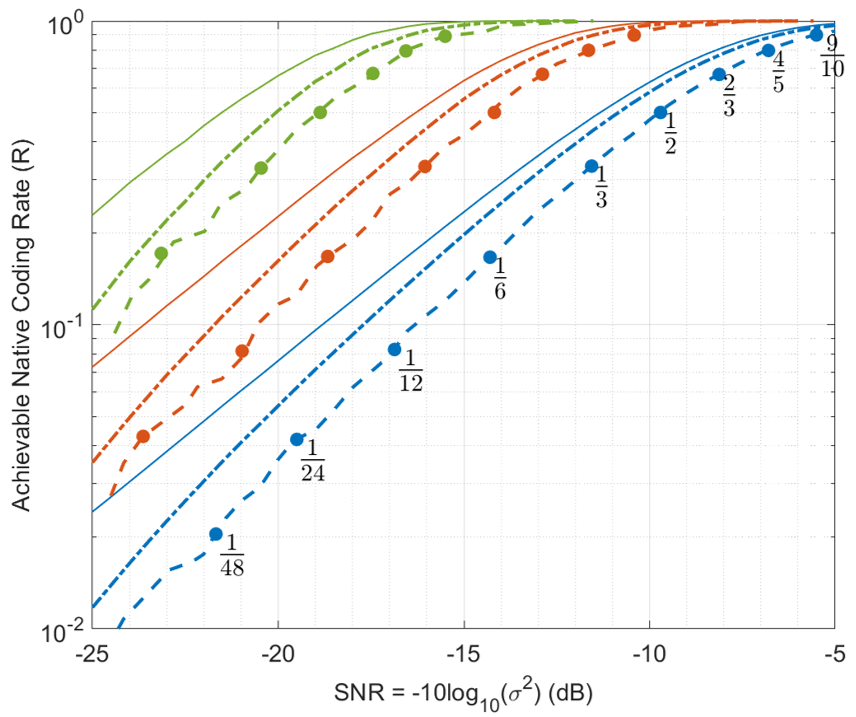}\label{fig:ccsk_achieve_native_rate_log}}\hfill%
\subfigure[Achievable effective coding rates]{\includegraphics[width=.33\linewidth]{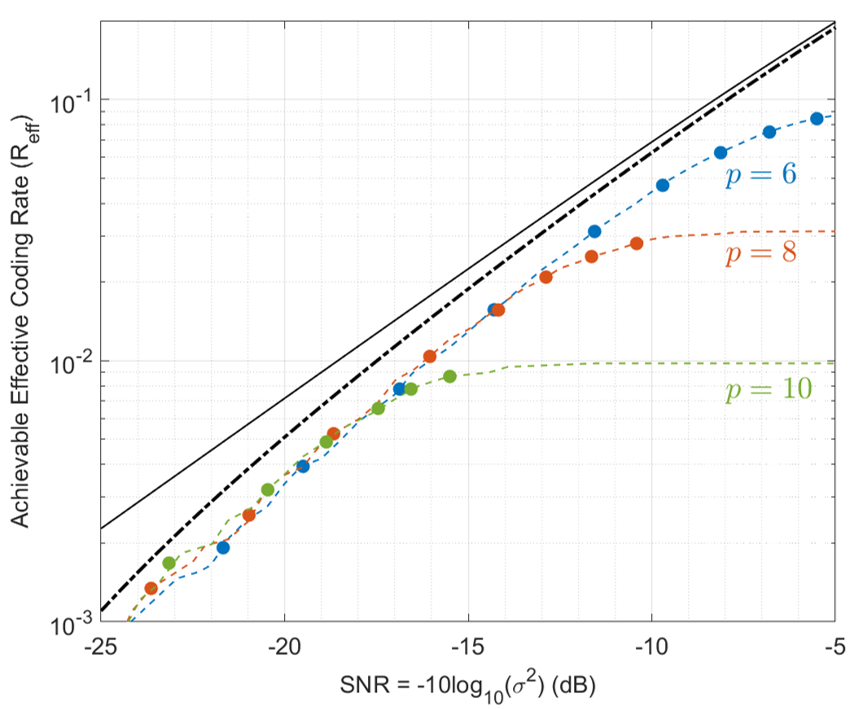}\label{fig:ccsk_achieve_effective_rate}}

\begin{minipage}{.74\linewidth}
\vspace*{5mm}\caption{Achievable native and effective coding rates for the AWGN \hfill Legend:\\ channel with CCSK modulated inputs, for a target $\text{WER}=10^{-4}$. }
\label{fig:ccsk_achieve_rates}
\end{minipage}\hfill%
\begin{minipage}{.25\linewidth}
\includegraphics[width=\linewidth]{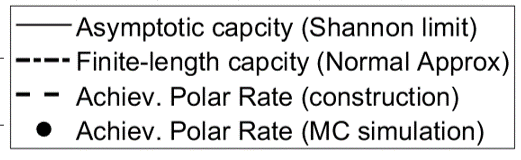}
\end{minipage}
\end{figure*} 

%

We assume that non-binary coded symbols in $\gf{q=2^p}$ are mapped into CCSK symbols of length $q$, which then undergo real-valued additive white Gaussian noise. The signal to noise ratio (SNR) value is defined as $\text{SNR} = -10\log_{10}(\sigma^2)$, where $\sigma^2$ is the noise variance. 
We have considered SNR values from $-25$ to $-5$\,dB, with a step of $0.5$\,dB, and for each SNR value we have constructed non-binary polar codes with parameters given in Table~\ref{tab:ccsk-nb-polar-parameters}. Recall that $n$ is the number of polarization steps, $N=2^n$ is the non-binary code length (number of coded symbols), $Np$ is the binary code length (number of coded bits), and $Nq$ is the effective number of transmitted bits (after CCSK modulation). 
The number of information bits, denoted $K_\text{bin}$, depends on the coding rate, and can be obtained by $K_\text{bin} = RNp$. We shall also refer to $R$ as the {\em native coding rate}, and define the {\em effective coding rate} $R_{eff} \defeq pR/q$, so that to take into account the spreading factor of the CCSK modulation.

\begin{table}[!t]
\centering
\caption{Parameters of the non-binary polar codes designed for the AWGN channel with CCSK modulated inputs}
\label{tab:ccsk-nb-polar-parameters}
\renewcommand{\arraystretch}{1.2}
\begin{tabular}{|c|c|c|c|c|}
\hline
$p$ & $n$ & $N$ & $Np$ & $Nq$ \\
\hline\hline
6 & 10 & 1024 & 6144 & 65536 \\
8 & 8  & 256  & 2048 & 65536 \\
10 & 6 & 64 &  640 & 65536 \\
\hline\hline
\end{tabular}
\end{table}

Fig.~\ref{fig:wer_polar_ccsk} shows the WER performance for various native coding rate values $R$, varying from $1/48$ to $9/10$. Two WER curves are shown for each native coding rate, a solid one, corresponding to Monte Carlo simulation results, and a dashed one, corresponding to the WER estimated at the code construction stage (Section~\ref{subsec:polar-code-construction}). It can be observed that the WER estimates we obtain at the code construction stage are tight.
Fig.~\ref{fig:ccsk_achieve_rates} shows the achievable native and effective coding rates, for a target $\text{WER}=10^{-4}$. The figure shows  the achievable coding rates for different Galois fields, obtained by using either the WER estimates at the code construction stage (dashed curves), or the WER obtained by Monte Carlo simulation (superimposed full markers). Moreover, dashed-dotted curves show the normal approximation of the maximum achievable rate in the finite block-length regime, while solid curves show the maximum achievable rate in the asymptotic block-length regime. It can be seen that the gap between the  achievable coding rates under non-binary polar coding and the normal approximation bound is about $1-1.5$\,dB. 

\section{Conclusion}
\label{sec:conclusion-polar} 

This paper investigated a new approach to reliable transmission of short data packets at very low signal-to-noise ratio, which combines CCSK modulation and non-binary polar coding. We proposed a design methodology for the non-binary polar code, aimed at accelerating the polarization speed, though maximizing the difference between the polarizing  parameters of the synthesized virtual channels. The proposed methodology is generic and may be used for other applications. Numerical results show that the system performance is close to the achievable limits in the finite blocklength regime. We expect that the observed performance may be further improved, by using a more powerful SC-List decoder~\cite{yuan2018construction}.

\bibliographystyle{IEEEtran}
\bibliography{biblio_polar}

\end{document}